\documentclass[a4paper]{article}

\usepackage{INTERSPEECH2021}
\usepackage{caption}
\usepackage{subcaption}
\usepackage{multicol}
\usepackage{multirow}

\title{Ctrl-P: Temporal Control of Prosodic Variation for Speech Synthesis}

\name{
    Devang S Ram Mohan$^1$\textsuperscript{*},
    Vivian Hu$^1$\textsuperscript{*},
    Tian Huey Teh$^1$\textsuperscript{*},
    Alexandra Torresquintero$^1$,
    Christopher G. R. Wallis$^1$,
    Marlene Staib$^1$,
    Lorenzo Foglianti$^1$,
    Jiameng Gao$^1$,
    Simon King$^{1,2}$}
\address{
  $^1$Papercup Technologies Ltd.,
  $^2$University of Edinburgh}
\email{\{devang,vivian,tian\}@papercup.com}

\usepackage{xspace}
\def\F0{$F_{0}$\xspace}

\begin{document}

\maketitle
\begingroup\renewcommand\thefootnote{*}
\footnotetext{These authors contributed equally to this work}
\endgroup
\begin{abstract}
Text does not fully specify the spoken form, so text-to-speech models must be able to learn from speech data that vary in ways not explained by the corresponding text. One way to reduce the amount of unexplained variation in training data is to provide acoustic information as an additional learning signal. When generating speech, modifying this acoustic information enables multiple distinct renditions of a text to be produced.

Since much of the unexplained variation is in the prosody, we propose a model that generates speech explicitly conditioned on the three primary acoustic correlates of prosody: \F0, energy and duration. The model is flexible about how the values of these features are specified: they can be externally provided, or predicted from text, or predicted then subsequently modified.

Compared to a model that employs a variational auto-encoder to learn unsupervised latent features, our model provides more interpretable, temporally-precise, and disentangled control. When automatically predicting the acoustic features from text, it generates speech that is more natural than that from a Tacotron 2 model with reference encoder. Subsequent human-in-the-loop modification of the predicted acoustic features can significantly further increase naturalness.
\end{abstract}
\noindent\textbf{Index Terms}: text-to-speech, controllable speech synthesis

\section{Introduction}

There are generally multiple ways in which a given text can be spoken. These distinct renditions may be the result of semantic distinctions, or different speaking styles, or simply natural random variation. In all cases, the differences are acoustic and are not fully specified by the text. Treating this variation as unwanted noise, and averaging it away, results in a lack of variation in the synthesised speech \cite{DBLP:journals/corr/abs-1905-07195}. `Average prosody' is probably meaningless and not the same as `default' prosody \cite{hodari2019ssw10}.

A popular approach to handling this unexplained variability is to learn a latent space \cite{DBLP:conf/icml/WangSZRBSXJRS18, DBLP:conf/iclr/HsuZWZWWCJCSNP19, DBLP:journals/corr/abs-1803-09047, DBLP:journals/corr/abs-1811-02122}. During inference, a sampled embedding from this latent space provides the information missing from the text. However, this approach can lead to undesirable artefacts in the synthetic speech \cite{henter2014measuring, henter2018deep} and, since the latent space is learned in an unsupervised fashion, it is difficult to choose an appropriate embedding to convey a specific desired rendition of the text. 

An alternative is to control specific acoustic features that correlate with prosodic variation. Since these can be automatically estimated from speech, it is simple to annotate the training data with their values. These acoustic features offer a direct way to synthesise prosodically distinct renditions of a text.

In our proposed model, we use three acoustic correlates of prosodic variation: $F_{0}$, energy and duration \cite{doi:10.1080/01690961003589492, prosodyofdubbedspeech}. Their values are specified per-phone in the force-aligned reference speech during training. Our modified version of the Tacotron 2 encoder-decoder model \cite{shenetaltaco2} attends over a concatenation of the encoder outputs and these acoustic features. The supervised nature of these features ensures stability of model training.

We also add to the model an acoustic feature predictor (AFP) to predict per-phone acoustic feature values, given the encoder outputs \cite{DBLP:conf/interspeech/ZengWCX20, raitio2020controllable}. This enables the model to produce natural synthesised speech from text alone, without requiring any additional inputs, whilst offering the option of control when desired.

The model thus contains interpretable, disentangled acoustic features that can be controlled at any desired temporal granularity, from individual phones to the whole utterance. This allows external human-in-the-loop control to generate multiple prosodically-distinct renditions of a given text.\footnote{\label{foot:samples}Samples: https://research.papercup.com/samples/temporal-control-interspeech-2021} Since the model predicts reasonable `default' values of these features from text, external control only needs to specify the subset of values to be changed.

\section{Related work}

Modelling prosodic variation has been an area of research interest for decades.
Unit-selection approaches include explicitly capturing variation in the recorded speech database \cite{DBLP:conf/interspeech/StromCK06}. Statistical parametric synthesisers used regression trees to map para-linguistic features to acoustic model parameters \cite{masuko2004style, nose2007style}. Given the ability of neural approaches to generate much more natural speech than these older systems \cite{shenetaltaco2, ren2020fastspeech, DBLP:journals/corr/abs-1710-07654}, recent research has focused on different ways of modelling prosodic variation within these approaches.

The model proposed is distinguished from preceding work by providing control over explicit acoustic features with good temporal precision. By `temporal precision' we mean both the ability to perform control at specified locations (e.g., per phone) and for those changes to result in localised changes in the generated speech.

\subsection{Explicit acoustic features vs. learnt latent dimensions}

A popular approach to incorporate prosodic variation into neural TTS systems is to treat it as a residual component \cite{DBLP:journals/corr/abs-1803-09047} -- that is, as acoustic variation not predictable from text -- and to learn a latent space which captures this information \cite{DBLP:conf/icml/WangSZRBSXJRS18, DBLP:conf/iclr/HsuZWZWWCJCSNP19, DBLP:journals/corr/abs-1803-09047, DBLP:journals/corr/abs-1811-02122, sun2020fullyhierarchical, klimkov2019finegrained}. Most commonly, the latent space is an embedding of an acoustic reference mel spectrogram, e.g., \cite{DBLP:conf/icml/WangSZRBSXJRS18, DBLP:journals/corr/abs-1803-09047, sun2020fullyhierarchical}.
Learnt latent spaces are generally uninterpretable and the dimensions are typically entangled. This renders the latent dimensions inconvenient for use as external control `levers'. Model estimation procedures to encourage disentanglement have been proposed \cite{sun2020fullyhierarchical}, but these appear to make the model highly sensitive to hyperparameter values, and hard to reproduce \cite{DBLP:journals/corr/abs-1811-12359}.

Instead of a learnt latent space, using acoustic features directly extracted from reference speech leads to stable and reproducible model estimation as well as ensuring interpretability. Our experimental results (Section \ref{sec:results}) demonstrate that the acoustic features are disentangled: they can be independently controlled, unlike the dimensions of the learnt latent space in a comparison model.

\subsection{Temporal precision}

The use of extracted acoustic features as additional input to a TTS model has recently been explored as a means of prosodic control, especially for \F0 \cite{ren2020fastspeech, morrison2020controllable, DBLP:journals/corr/abs-2006-06873}. In \cite{raitio2020controllable}, the authors demonstrate control over multiple extracted acoustic features, although by learning a \textit{global} embedding that is unable to provide control with temporal precision.

FastSpeech 2 \cite{ren2020fastspeech} is a non-autoregressive TTS model conditioned on extracted \F0 and energy features, which uses explicit phone durations. However, the features are required per-frame, which is not readily suitable for human-in-the-loop control. FastPitch \cite{DBLP:journals/corr/abs-2006-06873} addresses this challenge by modelling \F0 per-phone, but no longer models energy.

To the best of our knowledge, there is no previous model that provides precise and localised control over all of \F0, energy and duration at an appropriate temporal granularity that balances the competing requirements of i) accounting for unexplained acoustic variation, and ii) intuitive control for a human-in-the-loop.

\section{Proposed model and acoustic features}

Ctrl-P follows a multi-speaker Tacotron 2 \cite{shenetaltaco2, DBLP:conf/interspeech/ZhangWZWCSJRR19} attention-based encoder-decoder architecture, with modifications. A separately-trained WaveRNN vocoder \cite{kalchbrenner2018efficient} is used to generate a waveform from the mel spectrogram.

Let $p_{1}, ..., p_{N}$ denote the sequence of phones to be synthesised, and $\mathbf{y}_{1}, ..., \mathbf{y}_{T}$ denote the sequence of frames from the corresponding ground truth mel spectrogram. The proposed modification to the Tacotron 2 architecture consists of concatenating the sequence of encoder outputs $\mathbf{e}_{1}, ..., \mathbf{e}_{N}$ with the corresponding, phone-aligned acoustic features $\mathbf{a}_{1}, ..., \mathbf{a}_{N}$. In our experiments, the acoustic features are 3-dimensional. The decoder has access to this enhanced representation via the attention mechanism. We now describe how to obtain these acoustic features during training and inference.

\subsection{Training}

Using forced alignment, each phone $p_{i}$ aligns to a sequence of frames, $\mathbf{y}_{\alpha(i)}, ..., \mathbf{y}_{\beta(i)}$, from the ground truth mel spectrogram. The forced aligner uses the Kaldi Toolkit \cite{Povey_ASRU2011} with a model trained on the TTS training data.

\F0 is estimated using the RAPT algorithm \cite{talkin1995robust} and root mean square energy using the Librosa library \cite{brian_mcfee_2020_3606573}. We take the average of these features per-phone. The duration of phone $p_{i}$ is represented as the number of frames, $\beta_{i} - \alpha_{i} + 1$. For special tokens in the phone sequence, representing word and sentence boundaries, the value of all acoustic features is set to 0.

Each feature is normalised to zero mean and unit standard deviation, per speaker. These three features are then concatenated to form $\mathbf{a}_{i}$.

In initial experiments, taking the log of feature values \cite{raitio2020controllable, ren2020fastspeech} did not improve performance. Per-utterance normalisation also degraded performance. Neither were used in the experiments in Section \ref{sec:experimentalsetup}.

\subsection{Inference}

During inference, ground truth acoustic features are not required, but are predicted by the model. The acoustic feature predictor (AFP) takes as input $\mathbf{e}_{1}, ..., \mathbf{e}_{N}$ and predicts $\tilde{\mathbf{a}}_{1}, ..., \tilde{\mathbf{a}}_{N}$. The AFP consists of two stacked LSTM blocks, each comprising 2-layers of bidirectional LSTMs. The encoder outputs are mapped to a sequence of 64-dimensional hidden states by the first block, then to a sequence of 32-dimensional embeddings by the second block, then through a fully connected layer to a sequence of 16-dimensional embeddings and finally through a $\tanh$ non-linearity, followed by a projection to 3 dimensions to obtain the predicted acoustic features corresponding to each phone in the encoder input: $\tilde{\mathbf{a}}_{1}, ..., \tilde{\mathbf{a}}_{N}$.

\section{Experimental setup}
\label{sec:experimentalsetup}

We present experimental results to demonstrate that our proposed model, Ctrl-P, provides acoustic feature control that is more interpretable, disentangled and reproducible than the T-VAE benchmark. We also show that the proposed model generates more natural-sounding speech than the T-VAE or Tacotron-Ref benchmarks described in the next section, and that this naturalness can be further increased by human-in-the-loop control.

\subsection{Proposed model and benchmarks}

The data set used for all models was a proprietary, multi-speaker, Mexican-Spanish corpus consisting of approximately 38 hours of speech. Of this, $\sim$800 utterances were held-out for validation, with all speakers present in the training set being proportionately represented. We trained our modified Tacotron 2 model for 200k iterations using a weighted sum of gate loss and mel spectrogram reconstruction loss \cite{shenetaltaco2}. The model weights were then frozen and the AFP was trained for 400k iterations using an L1 loss between predicted and ground truth acoustic features $\mathbf{a}_{1}, ..., \mathbf{a}_{N}$. We used the Adam optimiser in both training phases.

In order to demonstrate the advantage of control using these explicit, extracted features, we compared the performance of our model against a temporal variational auto-encoder (henceforth: T-VAE) similar to the one described in \cite{DBLP:journals/corr/abs-1811-02122} which learns a latent space in an unsupervised fashion. This model uses a secondary attention mechanism to align an input reference mel spectrogram to the encoder outputs and thus produces a sequence of $3$-dimensional latents, one per encoder time step (i.e., per phone). A KL divergence loss is applied to the latent space to encourage its distribution to be close to a standard normal. During inference, these latents are predicted from the encoder output using a latent predictor (LP) whose architecture is identical to that of the AFP. The LP is trained in the same way as the AFP, using an L1 loss between the predicted latents and the latents produced from the reference mel spectrogram.

As a benchmark for naturalness, we used another well-established model: Tacotron 2 with a fixed-length global embedding predicted from a reference encoder \cite{DBLP:journals/corr/abs-1803-09047} (henceforth: Tacotron-Ref). Since this model offers no explicit control beyond providing a reference mel spectrogram, we did not include this in our evaluations of controllability.

\subsection{Waveform generation}

For the subjective naturalness evaluations, samples were vocoded with model-specific WaveRNN vocoders trained for 3M iterations on the mel spectrograms generated by their respective model for the training set. For the objective controllability evaluations, because subjective quality was not being measured, the large number of waveform samples required was generated using the Griffin-Lim algorithm \cite{griffinlim}.

\section{Results}
\label{sec:results}

\subsection{Disentangled control} \label{subsec:disentanglement}

We begin by demonstrating that our proposed model provides disentangled control over each individual acoustic feature. We modified the entire contour of each feature (or latent, in the case of the T-VAE benchmark) dimension by shifting it a fraction of the per-speaker standard deviation for that dimension. The other two dimensions were not shifted.

\begin{figure}
     \centering
     \begin{subfigure}[b]{0.527\columnwidth}
         \centering
         \includegraphics[width=\columnwidth]{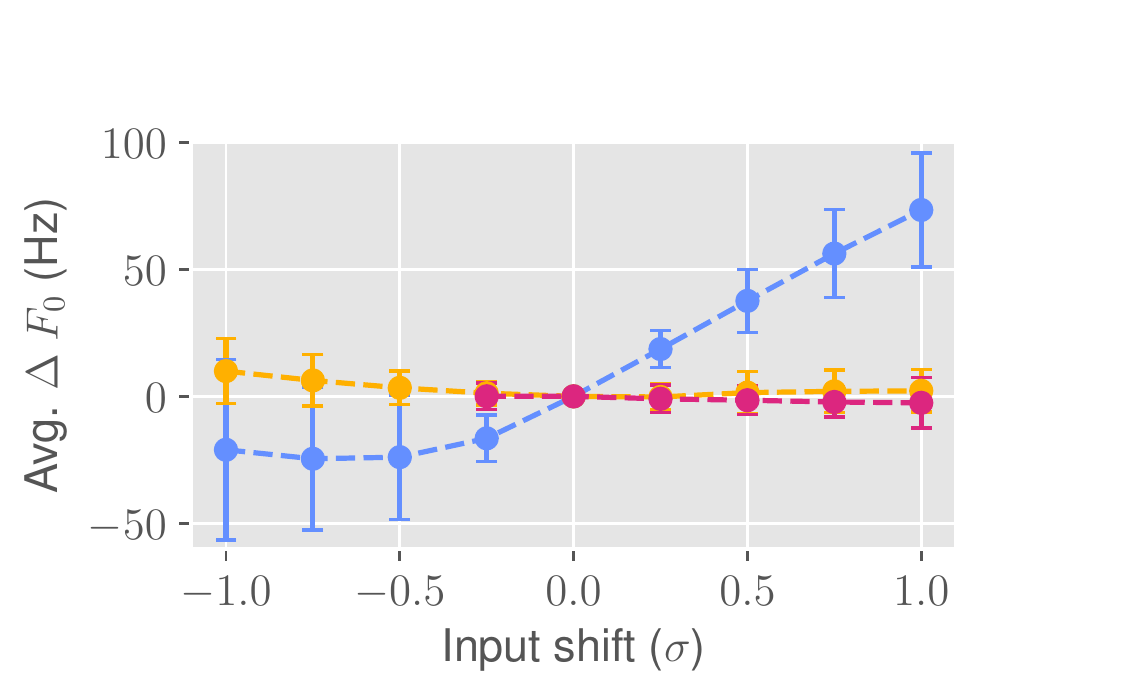}
         \caption{Ctrl-P control of \F0}
         \label{fig:disentanglement_tcess_pitch}
     \end{subfigure}%
     \begin{subfigure}[b]{0.473\columnwidth}
         \centering
         \includegraphics[width=\columnwidth]{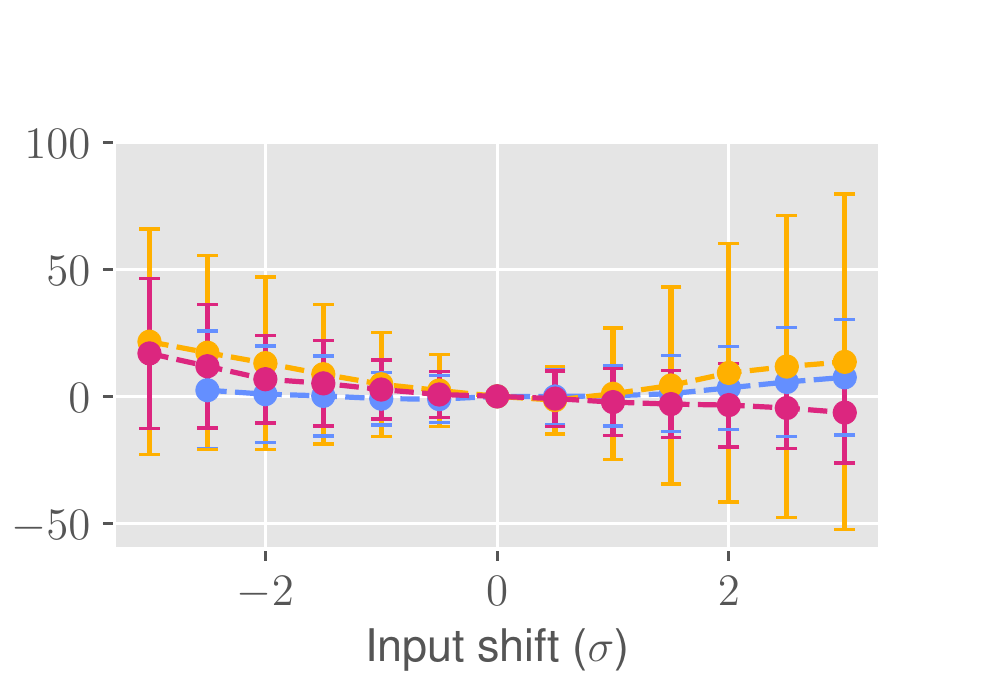}
         \caption{T-VAE control of \F0}
         \label{fig:disentanglement_tvae_pitch}
     \end{subfigure}
     
     \begin{subfigure}[b]{0.527\columnwidth}
         \centering
         \includegraphics[width=\columnwidth]{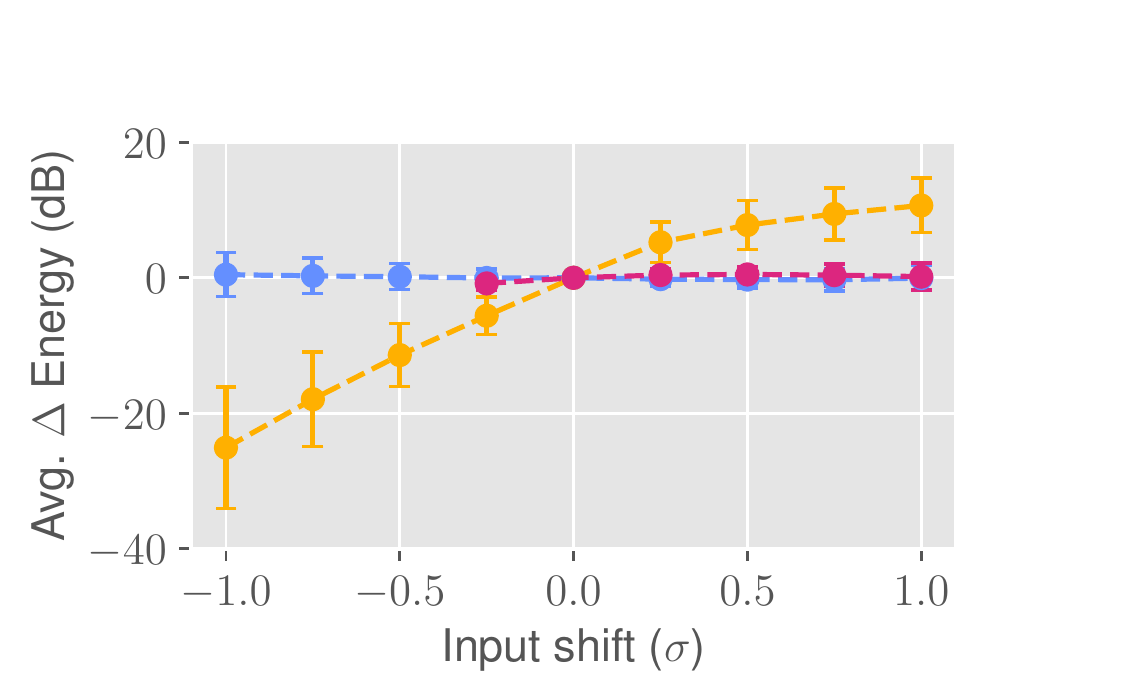}
         \caption{Ctrl-P control of energy}
         \label{fig:disentanglement_tcess_energy}
     \end{subfigure}%
     \begin{subfigure}[b]{0.473\columnwidth}
         \centering
         \includegraphics[width=\columnwidth]{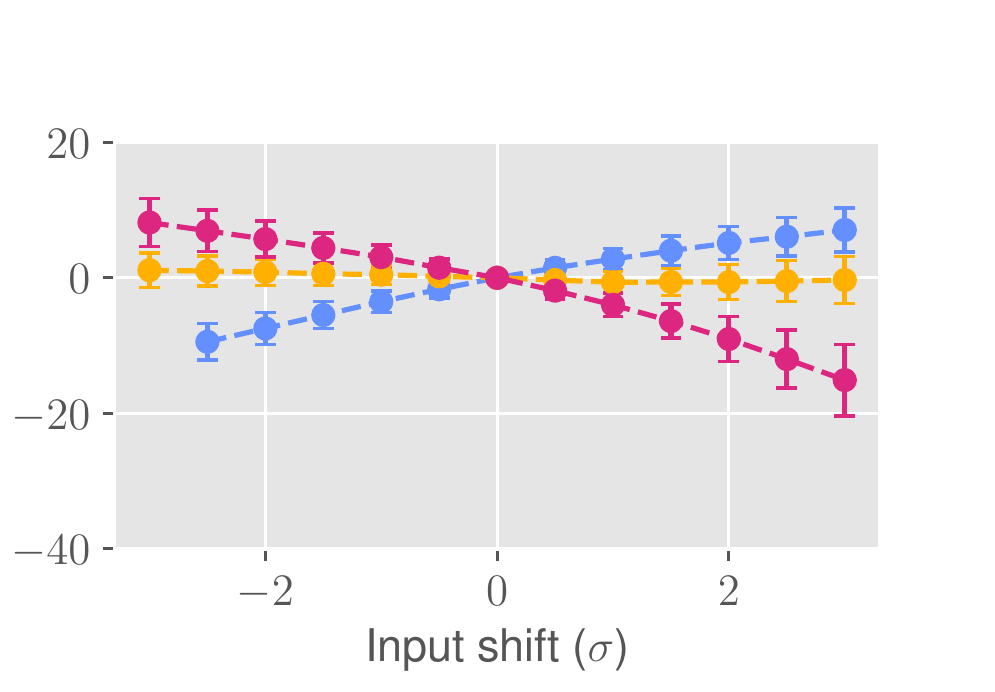}
         \caption{T-VAE control of energy}
         \label{fig:disentanglement_tvae_energy}
     \end{subfigure}

     \begin{subfigure}[b]{0.527\columnwidth}
         \centering
         \includegraphics[width=\columnwidth]{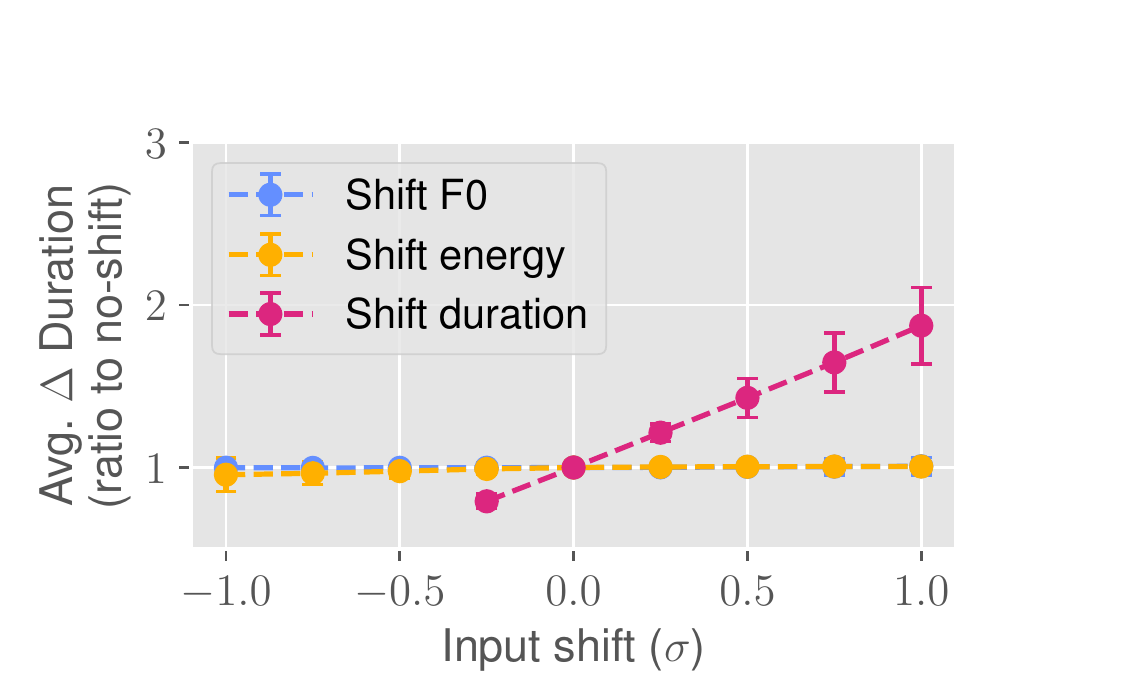}
         \caption{Ctrl-P control of duration}
         \label{fig:disentanglement_tcess_duration}
     \end{subfigure}%
     \begin{subfigure}[b]{0.473\columnwidth}
         \centering
         \includegraphics[width=\columnwidth]{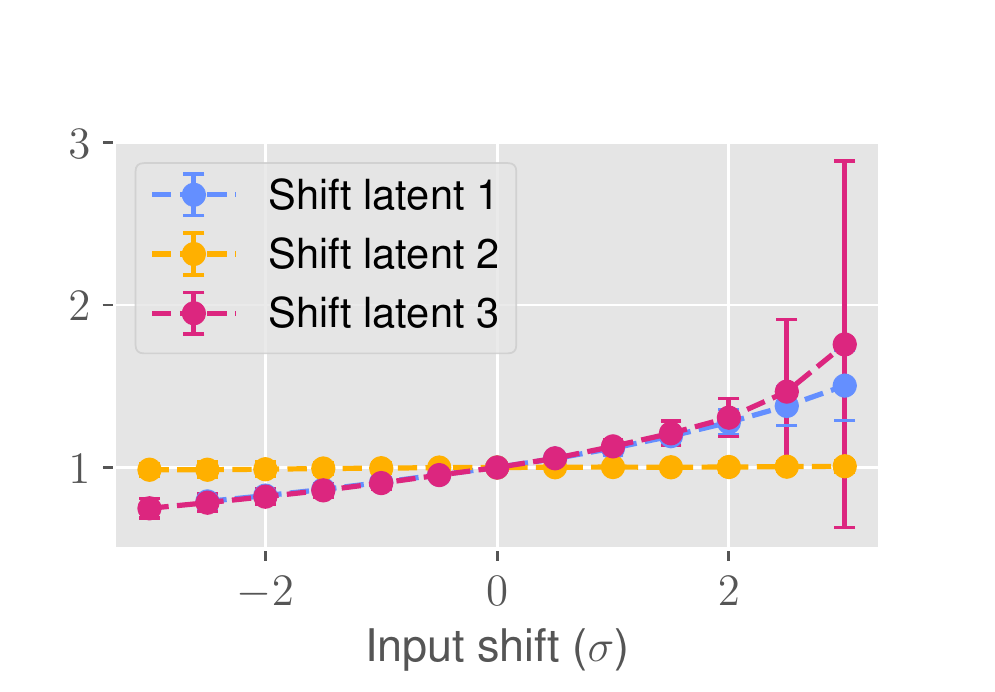}
         \caption{T-VAE control of duration}
         \label{fig:disentanglement_tvae_duration}
     \end{subfigure}
        \caption{Objective evaluation of disentangled control. x-axis: fraction of the speaker-specific standard deviation by which the feature (or latent) was shifted. Note that T-VAE required much larger changes to obtain comparable acoustic differences. y-axis: resulting change in that feature. Points represent the mean; whiskers denote one standard deviation. Shifting the duration below $-0.25 \sigma$ often generated speech that was too fast for the Kaldi aligner, so these data points are omitted. Results are averages across the validation set.}
        \label{fig:disentanglement}
\end{figure}

The average change in utterance-level \F0, energy and duration of the synthesised output was then measured. Figure \ref{fig:disentanglement} illustrates the measured changes for each acoustic feature. Figures \ref{fig:disentanglement_tcess_pitch}, \ref{fig:disentanglement_tcess_energy} and \ref{fig:disentanglement_tcess_duration} show that the Ctrl-P model is able to make changes only to the specified feature. For example, increasing the \F0 feature results only in an increase in \F0 of the synthesised output. In contrast, T-VAE produces entangled changes (Figures \ref{fig:disentanglement_tvae_pitch}, \ref{fig:disentanglement_tvae_energy} and \ref{fig:disentanglement_tvae_duration}). Not only does a change in the value of an individual latent result in a hard-to-interpret change of multiple acoustic properties, those changes can be inconsistent across utterances as seen in the wide standard deviation bands (e.g., changes in \F0 from shifting T-VAE latent 2).

\subsection{Temporally-precise control}

We randomly selected a subset of the stressed vowels within an utterance and shifted each feature (or latent) dimension in turn for that phone, leaving other phones unchanged. Forced alignment was used to label the modified phones in the synthetic output waveform and the change in \F0, energy and duration was measured.

\begin{figure}
     \centering
     \begin{subfigure}[b]{0.527\columnwidth}
         \centering
         \includegraphics[width=\columnwidth]{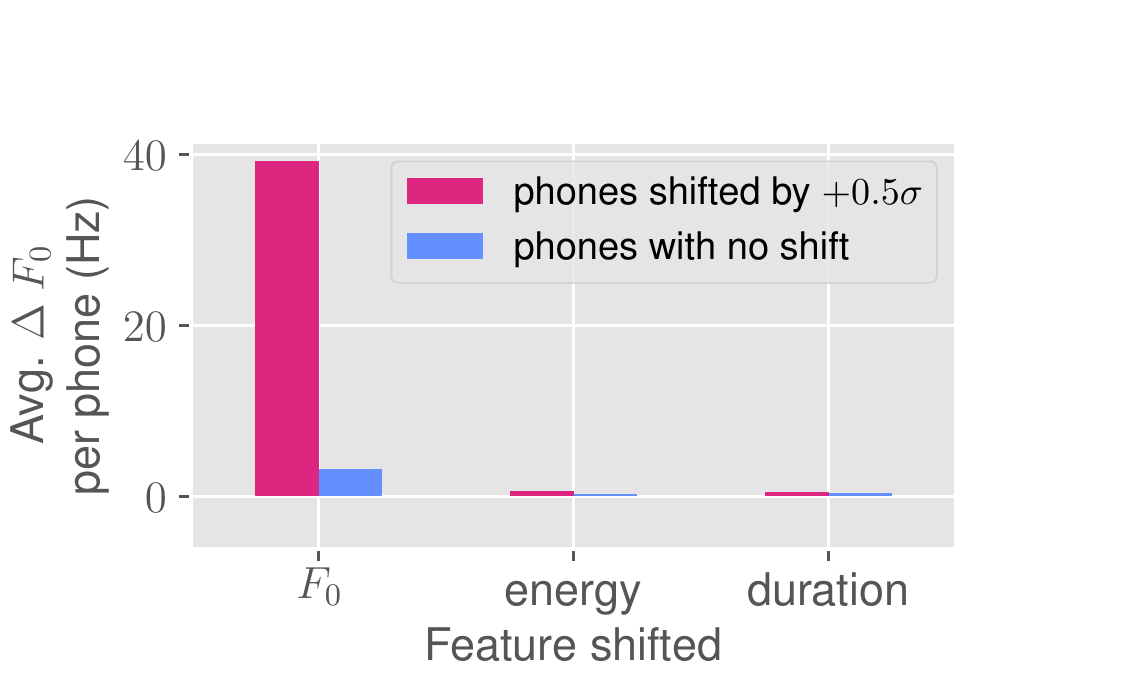}
         \caption{Ctrl-P control of \F0}
         \label{fig:temporal_tcess_pitch}
     \end{subfigure}%
     \begin{subfigure}[b]{0.473\columnwidth}
         \centering
         \includegraphics[width=\columnwidth]{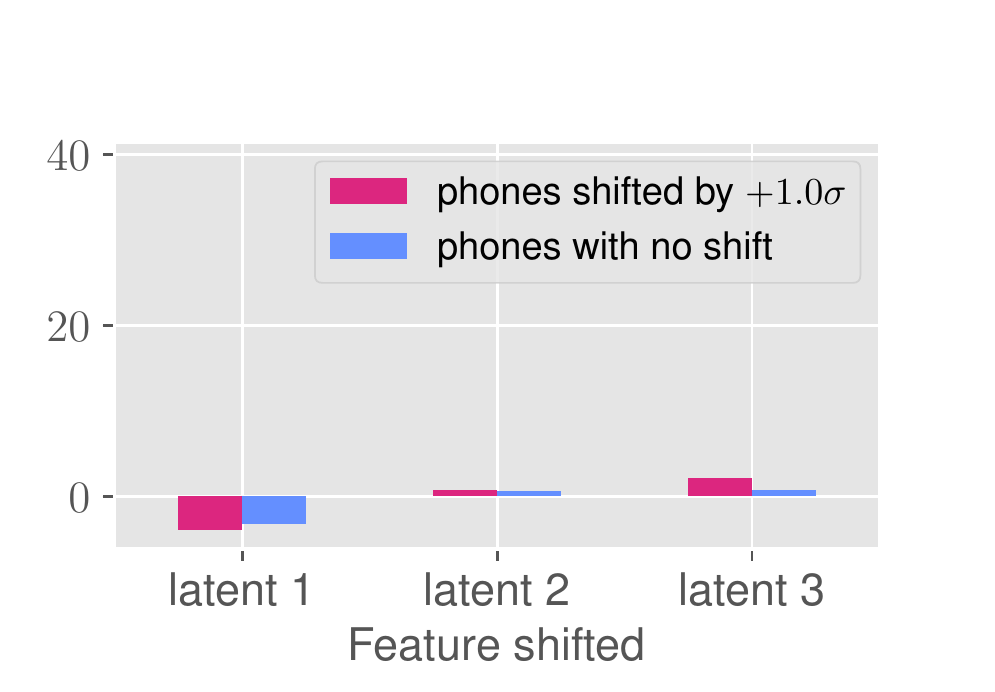}
         \caption{T-VAE control of \F0}
         \label{fig:temporal_tvae_energy}
     \end{subfigure}
     
     \begin{subfigure}[b]{0.527\columnwidth}
         \centering
         \includegraphics[width=\columnwidth]{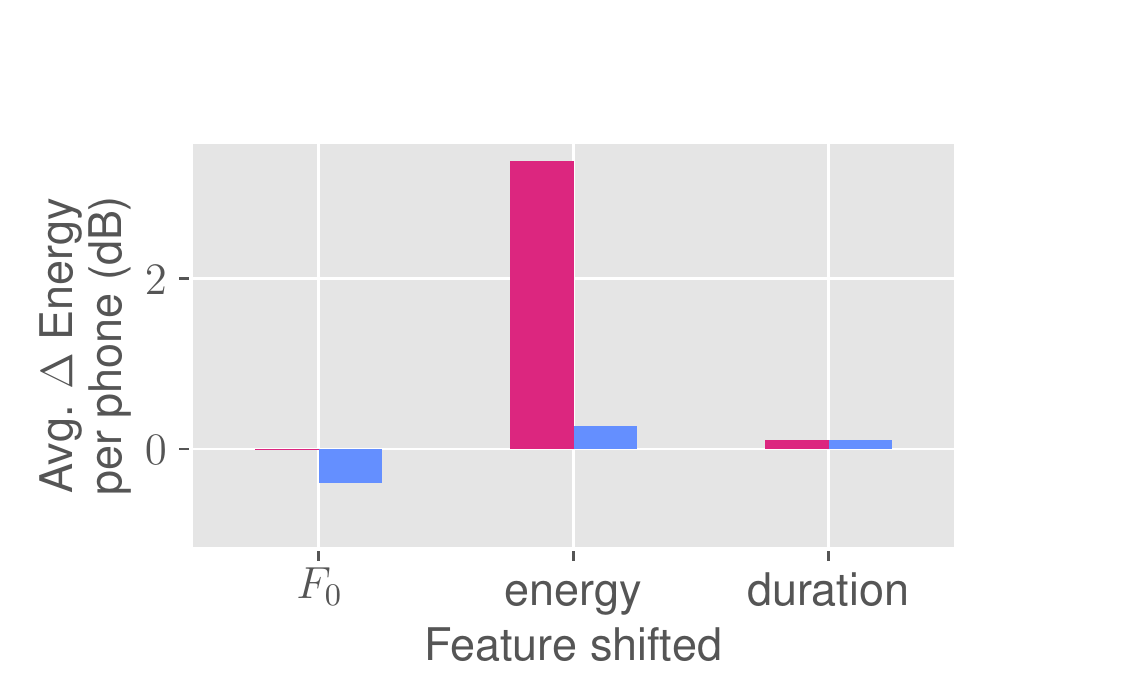}
         \caption{Ctrl-P control of energy}
         \label{fig:temporal_tcess_duration}
     \end{subfigure}%
     \begin{subfigure}[b]{0.473\columnwidth}
         \centering
         \includegraphics[width=\columnwidth]{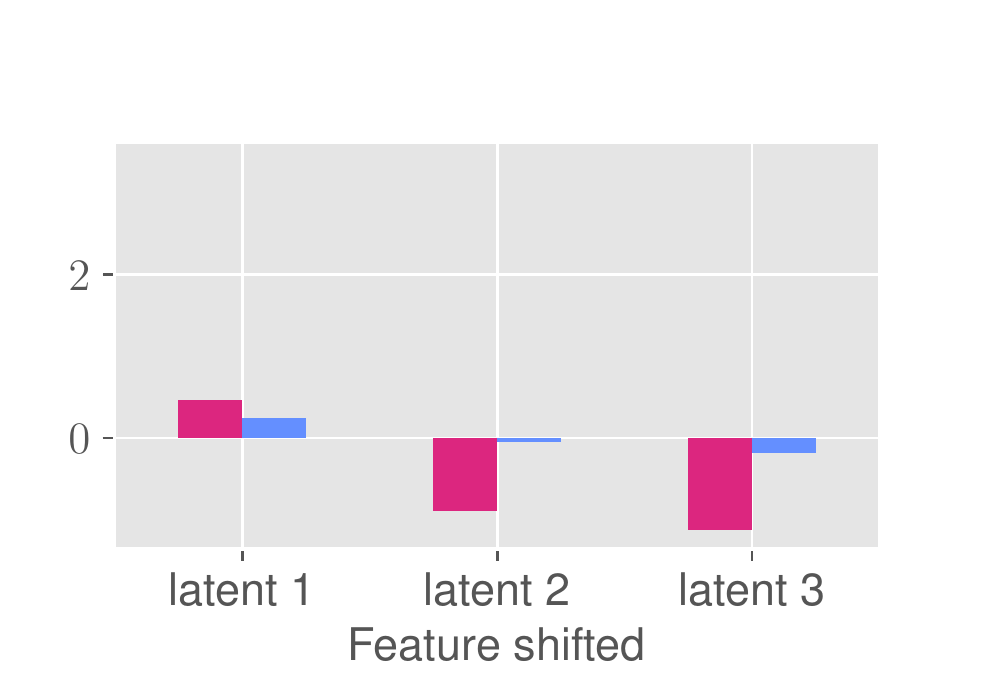}
         \caption{T-VAE control of energy}
         \label{fig:temporalt_tvae_pitch}
     \end{subfigure}

     \begin{subfigure}[b]{0.527\columnwidth}
         \centering
         \includegraphics[width=\columnwidth]{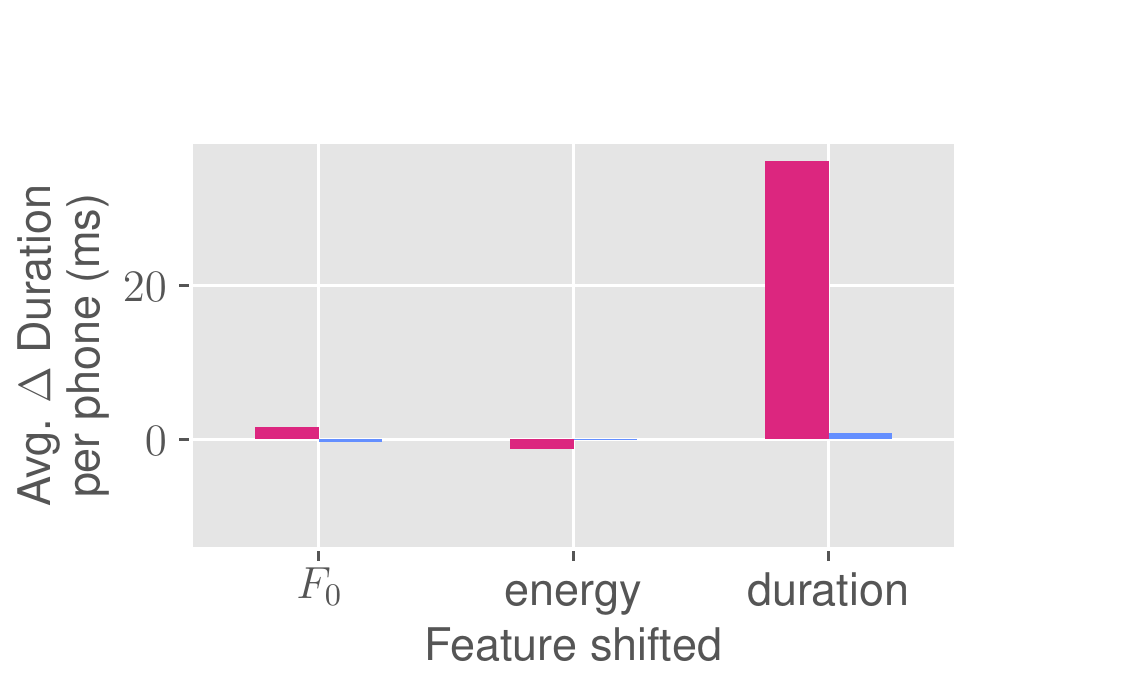}
         \caption{Ctrl-P control of duration}
         \label{fig:temporal_tcess_energy}
     \end{subfigure}%
     \begin{subfigure}[b]{0.473\columnwidth}
         \centering
         \includegraphics[width=\columnwidth]{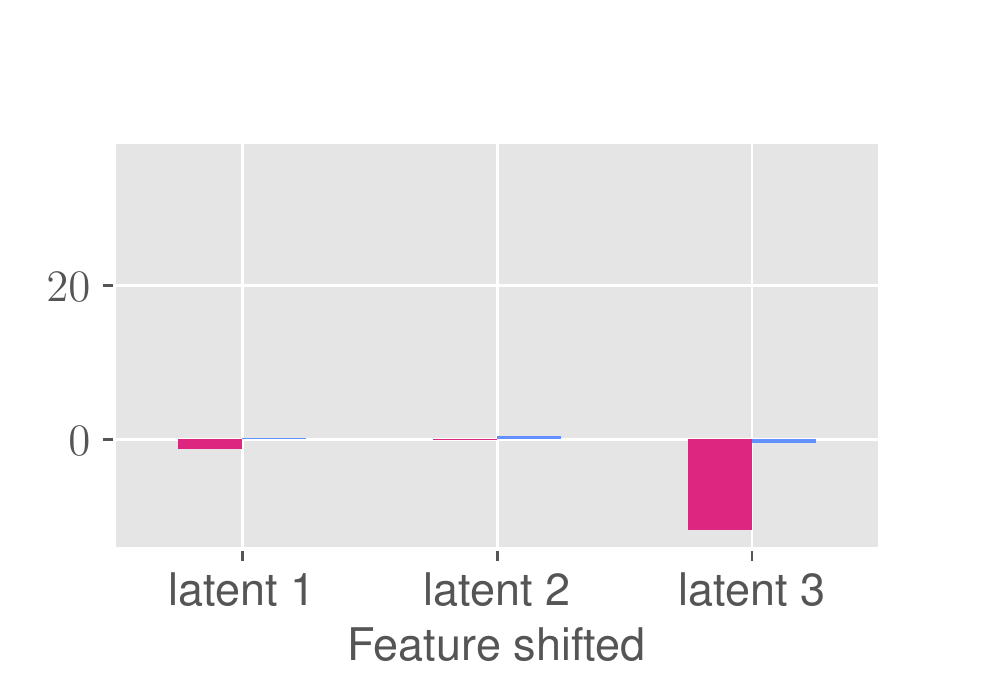}
         \caption{T-VAE control of duration}
         \label{fig:temporal_tvae_duration}
     \end{subfigure}
        \caption{Objective evaluation for precision of temporal control. Results are averaged across all 65 validation set utterances generated for a randomly chosen female speaker. Similar behaviour was observed for all speakers.}
        \label{fig:temporal_control}
\end{figure}

Figure \ref{fig:temporal_control} shows that Ctrl-P achieves temporally-precise and disentangled control of only the intended phones, for all three features. In contrast, the temporal region of influence is unclear for T-VAE, with both the modified and unmodified phones undergoing changes in acoustic properties. This might be attributed to the use of an attention mechanism to align latents with phones in T-VAE, in contrast to the hard alignment used by Ctrl-P.

To illustrate the utility of having this fine-grained level of control, we provide samples with varying renditions of the same text that result in semantically distinct utterances.\footnotemark[1]

\subsection{Reproducibility}
\label{section:results:reproducibility}

\begin{figure}
     \centering
     \begin{subfigure}[b]{0.527\linewidth}
         \centering
         \includegraphics[width=\linewidth]{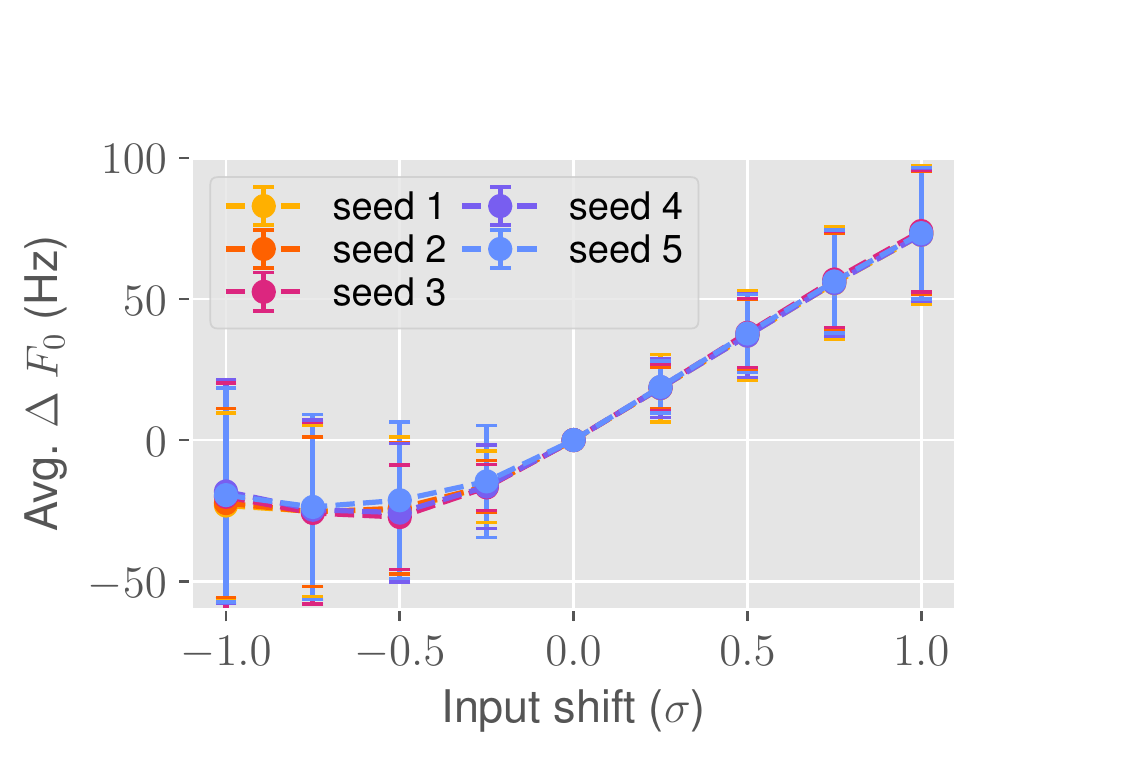}
         \caption{Ctrl-P changing \F0}
         \label{fig:robustness_tcess_pitch}
     \end{subfigure}%
     \begin{subfigure}[b]{0.473\linewidth}
         \centering
         \includegraphics[width=\linewidth]{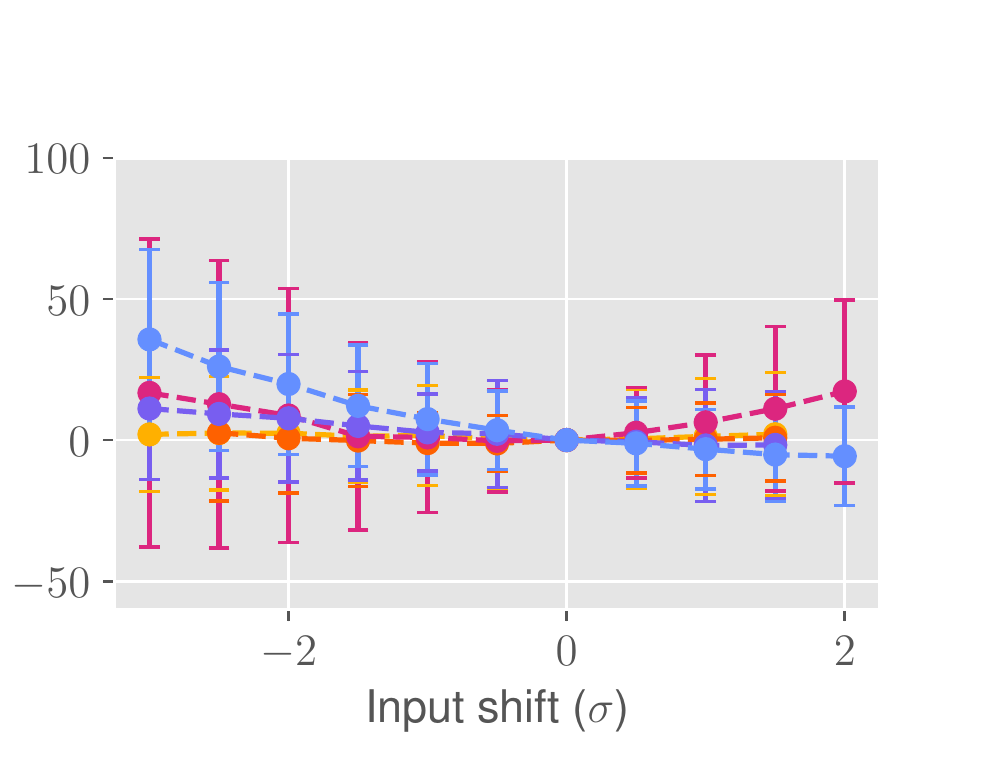}
         \caption{T-VAE changing latent 1}
         \label{fig:robustness_tvae_pitch}
     \end{subfigure}
     \begin{subfigure}[b]{0.527\linewidth}
         \centering
         \includegraphics[width=\linewidth]{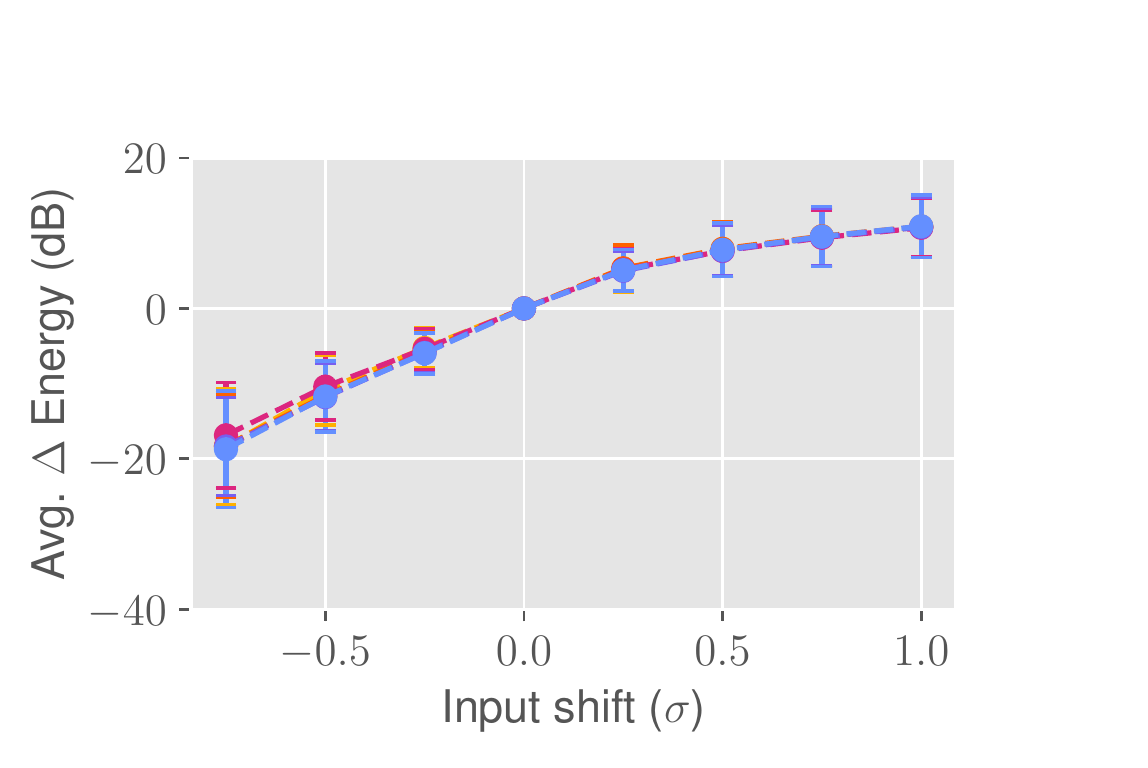}
         \caption{Ctrl-P changing energy}
         \label{fig:robustness_tcess_energy}
     \end{subfigure}%
     \begin{subfigure}[b]{0.473\linewidth}
         \centering
         \includegraphics[width=\linewidth]{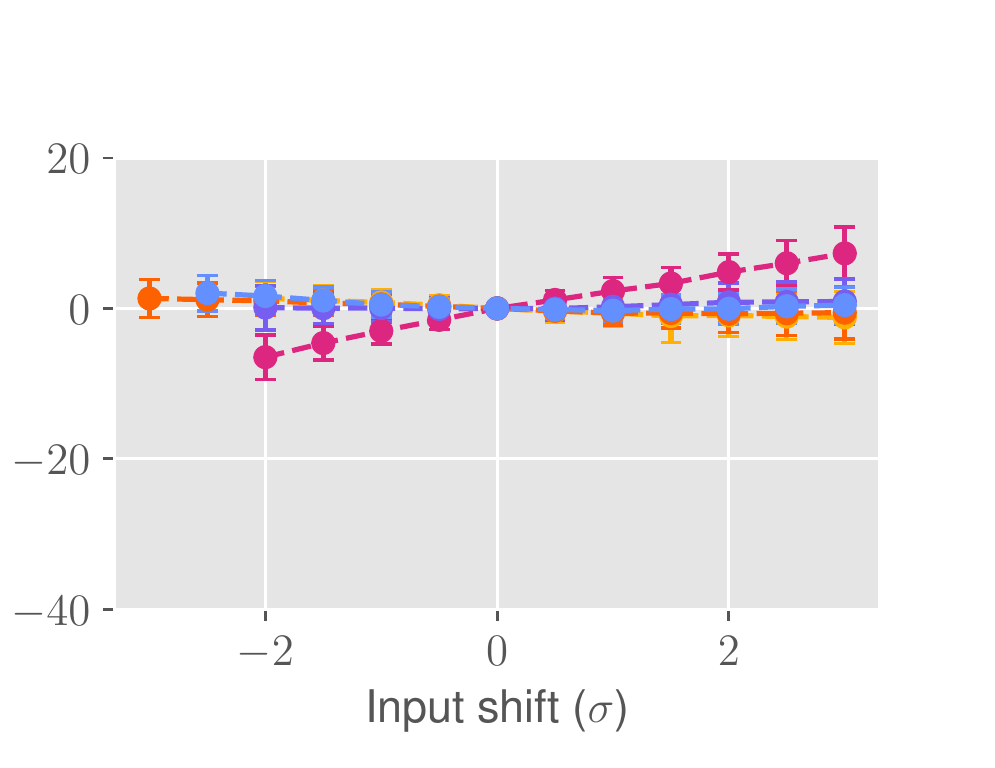}
         \caption{T-VAE changing latent 2}
         \label{fig:robustness_tvae_energy}
     \end{subfigure}     
     \begin{subfigure}[b]{0.527\linewidth}
         \centering
         \includegraphics[width=\linewidth]{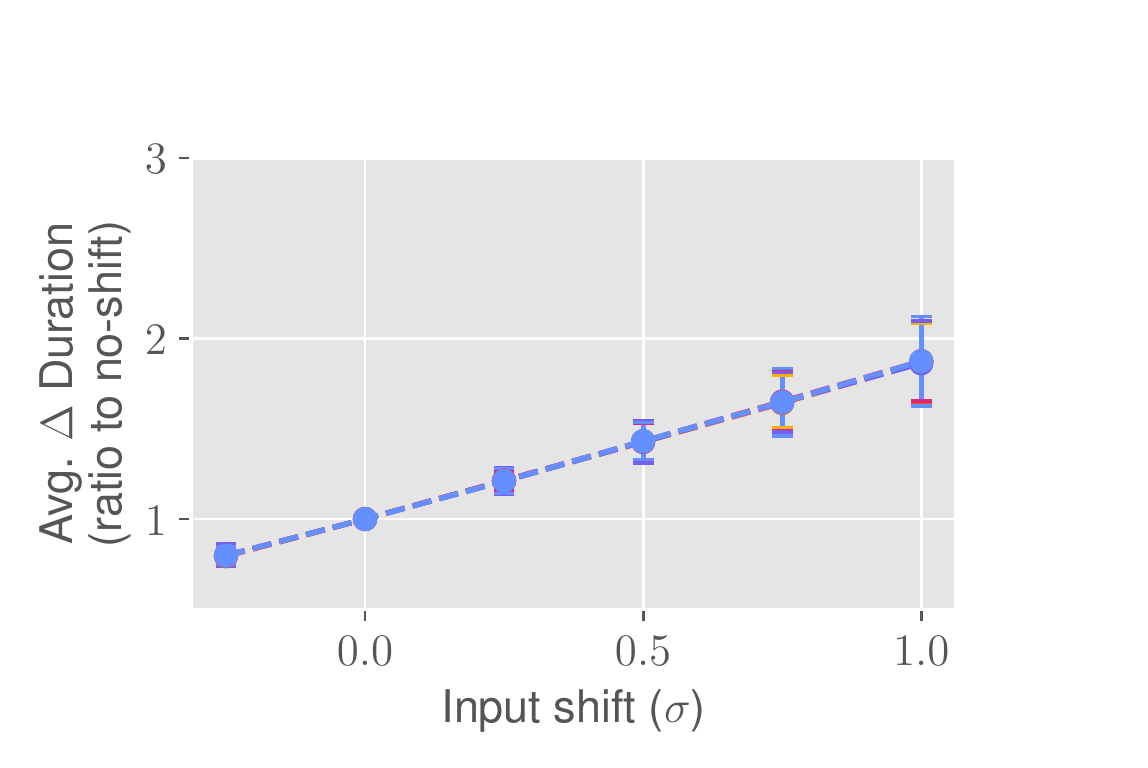}
         \caption{Ctrl-P changing duration}
         \label{fig:robustness_tcess_duration}
     \end{subfigure}%
     \begin{subfigure}[b]{0.473\linewidth}
         \centering
         \includegraphics[width=\linewidth]{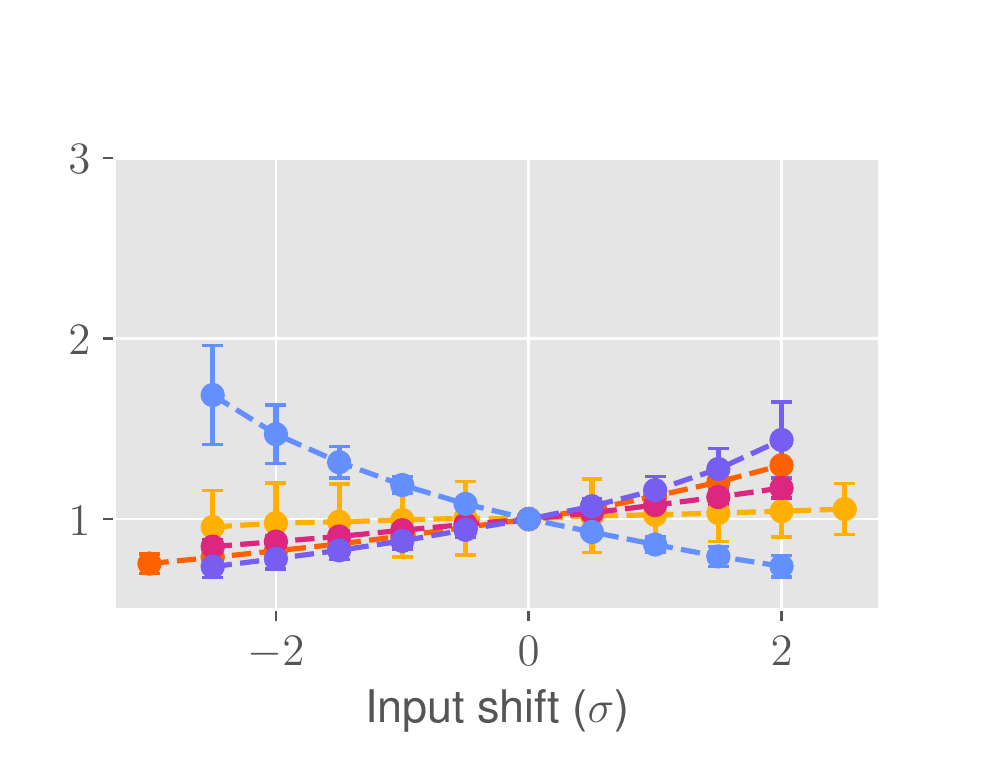}
         \caption{T-VAE changing  latent 3}
         \label{fig:robustness_tvae_duration}
     \end{subfigure}         
        \caption{Objective evaluation of model reproducibility. Shifting each input acoustic feature in Ctrl-P results in predictable changes in the generated speech across random seeds. Shifting each latent dimension in T-VAE results in unpredictable changes.}
        \label{fig:robustness}
\end{figure}

By retraining Ctrl-P and T-VAE from different random seeds, then applying the same analysis as Section \ref{subsec:disentanglement}, we are able to quantify the reproducibility of each model. Figure \ref{fig:robustness} shows that the effect of the T-VAE latent dimensions varies substantially across random seeds. There is no such sensitivity for Ctrl-P.

\subsection{Naturalness}

\begin{figure}
     \centering
     \includegraphics[width=\linewidth]{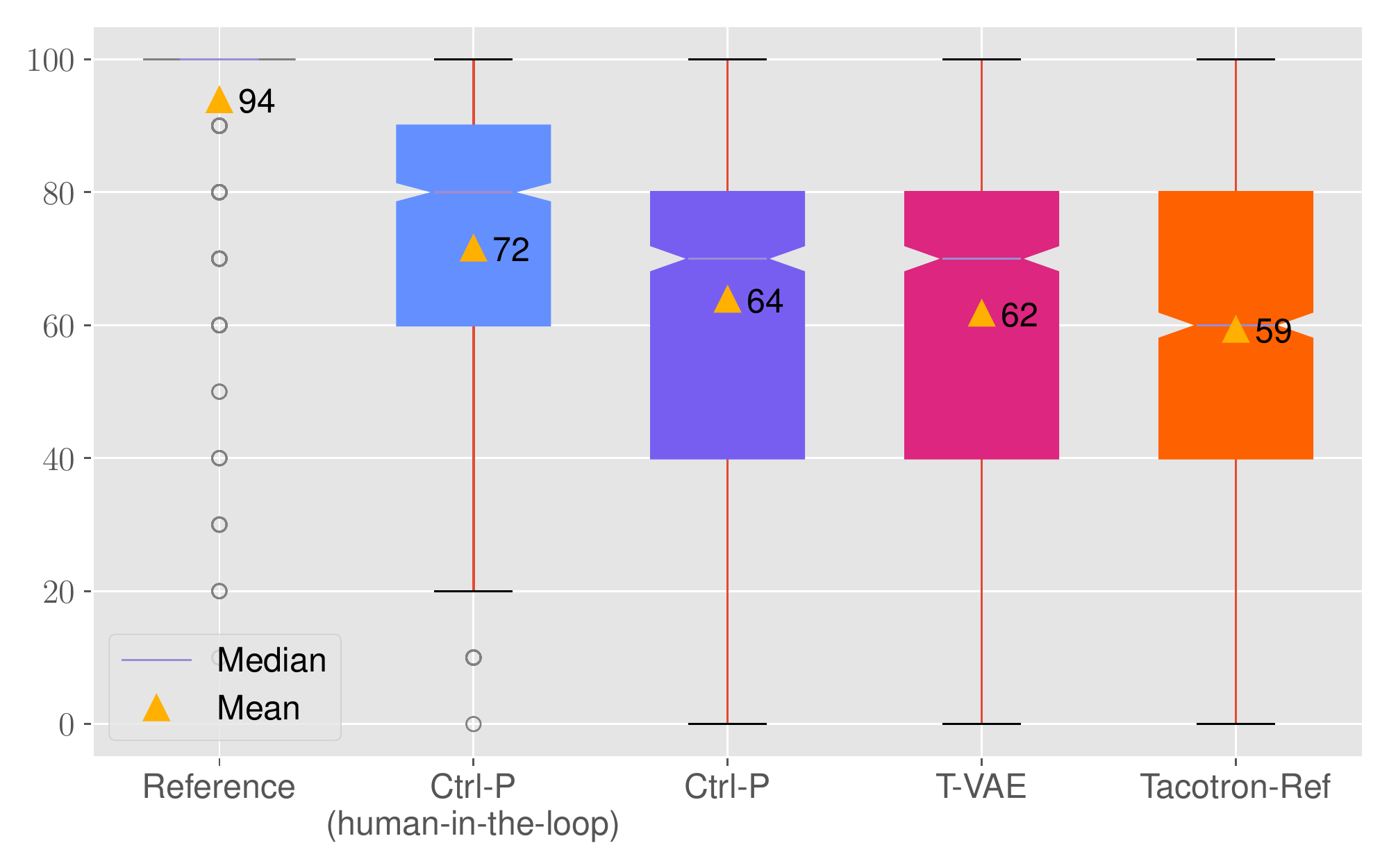}
     \caption{Results from a MUSHRA-like subjective evaluation of naturalness. Each box spans the 1st to 3rd quartiles (Q1, Q3); whiskers denote the range (capped at $1.5 \times (Q3-Q1)$); outliers are shown as individual points. }
     \label{fig:naturalness_mushra}
\end{figure}

Figure \ref{fig:naturalness_mushra} presents the results from a MUSHRA-like listening test \cite{mushraref} of a randomly chosen selection of 5 validation utterances each from 6 speakers (3 male and 3 female). The samples for Ctrl-P and T-VAE were generated using standard inference (i.e., acoustic features or latents were automatically predicted and not modified). For inference with Tacotron-Ref, the speaker-specific mean embedding estimated from the training set is used.

To create the Ctrl-P (human-in-the-loop) samples, the acoustic features predicted by the AFP were modified by a human, aiming for higher fidelity to the original reference. The uninterpretable and entangled behaviour of the T-VAE latents made such human control impractical for T-VAE. Samples from these 4 models plus a hidden reference (natural speech) and anchor were presented to 50 Spanish-speaking listeners recruited via Amazon Mechanical Turk. 

Listeners were asked to rate the naturalness of each sample on a 0-100 scale in intervals of 10, based on how similar to the reference they sound.

We filtered out listeners who failed to identify the hidden reference (by ranking it the highest) more than 50\% of the time, leaving 32 valid listeners. Results are presented in Figure \ref{fig:naturalness_mushra}; all pairs significantly differ in naturalness (two-sided t-test with Holm-Bonferroni correction; $p \leq 0.05$).

We observe that the standard Ctrl-P model is able to generate marginally more natural speech than the T-VAE model. This is despite Ctrl-P being constrained to encoding the supervised acoustic features whilst T-VAE is able to encode \textit{any} information it chooses from the reference mel spectrogram. Both models were found to produce more natural speech than Tacotron-Ref.

Human-in-the-loop control of the Ctrl-P acoustic features resulted in a further increase in naturalness. The temporal precision offered by the model enabled annotators to make specific, targeted adjustments to the rhythm, intonation and word emphasis within the utterance. Moreover, this modification of the acoustic features did not negatively impact the ability of the matched neural vocoder (trained for Ctrl-P) to generate high-quality waveforms.

\section{Conclusions and future work}

By modelling \F0, energy and duration explicitly, the proposed model provides interpretable, disentangled, and temporally-precise control over those properties in the generated speech. Model training is reproducible since it is not overly-sensitive to random seed. The chosen feature set could be expanded to include other acoustic correlates of prosody, such as spectral tilt or segmental reduction.

Future work might focus on improving the feature predictions from the AFP by exploring alternate architectures and training routines to obtain improved `default' prosody for the model, in the absence of, or as a better starting point for, human-in-the-loop control. We also observed a tapering of the influence of control at the extremes of feature values seen during training. Additional research could address this, aiming for a model that is able to generalise \textit{beyond} the range of feature values found in the data.

Our results demonstrate that the model is amenable to human-in-the-loop modifications of the synthetic speech. However, whilst per-phone control over the three principal acoustic correlates of prosody enabled improvements in naturalness to be achieved, it may be preferable to provide more abstract controls such as `emphasise this word', or `create rising question intonation'.

\vspace{2mm}
\noindent\textbf{Acknowledgements} We thank our adviser Mark Gales for feedback on this work.

\newpage
\bibliographystyle{IEEEtran}

\bibliography{template}

\end{document}